\documentclass[twocolumn,aps,showpacs,showkeys,prd,superscriptaddress,byrevtex,longbibliography]{revtex4-1}

\usepackage{amsmath,latexsym}
\usepackage{xcolor}
\usepackage{etoolbox}
\usepackage{breqn}

\usepackage[colorlinks=true,urlcolor=blue,linkcolor=blue,citecolor=blue]{hyperref}

\makeatletter
\let\cat@comma@active\@empty
\makeatother

\usepackage{graphicx}

\begin{document}
\preprint{OU-HET-1045}

\title{Universal Higher-Order Topology from 5D Weyl semimetal: \\
Edge topology, edge Hamiltonian and nested Wilson loop}
\author{Koji Hashimoto}
\email{koji@phys.sci.osaka-u.ac.jp}
\noaffiliation
\author{Yoshinori Matsuo}
\email{matsuo@het.phys.sci.osaka-u.ac.jp}
\noaffiliation
\affiliation{Department of Physics, Osaka University, Toyonaka, Osaka 560-0043, Japan}

\begin{abstract}
Higher-order topological insulators (HOTIs) or multipole insulators,
hosting peculiar corner states,
were discovered \cite{benalcazar2017quantized,schindler2018higher}.
It was independently discovered
\cite{hashimoto2017edge} that continuum 5-dimensional (5D) Weyl semimetals generically host the corner states,
and so do 4D class A and 3D class AIII topological insulators.
In this paper we further confirm that the 5D Weyl semimetals, upon dimensional reduction,
lead to universal higher-order topology.
First we explain a discrete symmetry protecting the 5D Weyl semimetals, and 
describe dimensional reductions of the 5D Weyl semimetals to the popular HOTIs in the continuum limit.
We calculate the topological charge carried by edge states of the 5D Weyl semimetal,
for the most generic boundary condition. 
The topological charge is a Dirac monopole,
which can also be seen from that 
edge Hamiltonians are always of the form of a 3D Weyl semimetal. 
This edge topology leads to the edge-of-edge states, or the corner states, generically, 
suggesting that the 5D Weyl semimetal is thought of as a physical structural origin of corner states
in HOTIs.
In addition, we explicitly calculate a nested Wilson loop of the 5D Weyl semimetal
and find that the topological structure is identical to that of a Wilson loop of a Dirac monopole.
\end{abstract}


\maketitle

\setcounter{footnote}{0}

\noindent

\section{Introduction}

Among various topological phases of matter, recently higher-order topology in condensed matter physics 
attracted attention, due to its peculiarity in dimensional aspects. Gapless states can show up not on the
surface of materials but on corners or hinges, thus localized states as a result of the bulk topology
has a spatial support whose dimensions are lower by two or more, compared to the bulk dimensions.
The higher-order topology goes beyond the ordinary bulk-edge correspondence
\cite{jackiw1976solitons,hatsugai1993chern,wen2004quantum}, thus any universal theoretical understanding
of the higher-order topology is in demand.

Higher order topological insulators (HOTI) were introduced in seminal papers
\cite{benalcazar2017quantized,schindler2018higher}
(see for earlier related proposals \cite{volovik2010topological,slager2015impurity}, and a mathematical proposal
\cite{hayashi2018topological}). The notion and mechanism of corner states were
independently introduced in \cite{hashimoto2017edge} (see also \cite{hashimoto_edge_of_edge}), 
where a 5D Weyl semimetal was shown to
host the corner states quite generically. 

The Weyl semimetals \cite{armitage2018weyl} have particularly simple
topological structures, 
and their higher dimensional generalization and its edge states were studied \cite{lian2016five}.
It was discovered in \cite{hashimoto2016topological} that an edge state of the continuum 5D Weyl semimetal
has a nontrivial topological charge, through an analogy of Nahm construction of monopoles \cite{nahm1980simple}
and a superstring T-duality. This observation lead to the notion of corner states in \cite{hashimoto2017edge}, since the
topological charge of the edge state means the existence of a corner state. The corner state was called in \cite{hashimoto2017edge} as an ``edge-of-edge state."

In this paper we bridge the description of the corner states obtained in the 5D Weyl semimetals \cite{hashimoto2017edge} and
that of the popular HOTIs \cite{benalcazar2017quantized,schindler2018higher}. 
We first show that the popular HOTIs in the continuum are obtained by dimensional reduction of the 5D Weyl semimetal,
and identify the discrete symmetry of the 5D Weyl semimetals.
We explicitly describe the edge Hamiltonian and the edge topological charge
of the 5D Weyl semimetal with a completely generic boundary condition, to show that the edge hosts
a nontrivial topology of the form of a Dirac monopole. As for the generic boundary condition we 
follow the strategy developed in \cite{hashimoto2017boundary} for continuum Weyl semimetals.
As HOTIs are often characterized by nested Wilson loops \cite{benalcazar2017quantized} and 
entanglement polarization \cite{fukui2018entanglement}, we explicitly calculate the nested Wilson loop 
of the 5D Weyl semimetals, to confirm that it is governed by a topology of a Dirac monopole.
Therefore, in total, the higher order topology of the 5D Weyl semimetals is due to the common Dirac monopole
structure 
in all of the following aspects: (1) edge Hamiltonian, (2) topology of the edge state, and (3) nested Wilson loop Hamiltonian.

This topological structure governed by the 5D Weyl semimetals can be identified in various recent study of the 
higher-order topology. For example,
topological quadrupole semimetals \cite{lin2018topological} hosts a topological semimetal structure
on their surfaces, and the 5D Weyl semimetal shares \cite{hashimoto2017edge} the same surface semimetal structure
\footnote{The higher-order topological semimetal (2) of \cite{lin2018topological} has the same 
structure as the system given in Sec.~V.A 
of \cite{hashimoto2017edge} with a reduction $p_1=0$ and $p_5=m$.}.
From these observations, 
we claim that, following \cite{hashimoto2017edge}, the 5D Weyl semimetal is a universal origin of the higher-order topology.
Namely, for a class of microscopic Hamiltonians whose continuum limit is that of the continuum 5D Weyl semimetal
(or a dimensional reduction of it), we claim that universally an edge-of-edge state appears.

\begin{figure*}[t]
	\begin{center}
	 \includegraphics[width=50em]{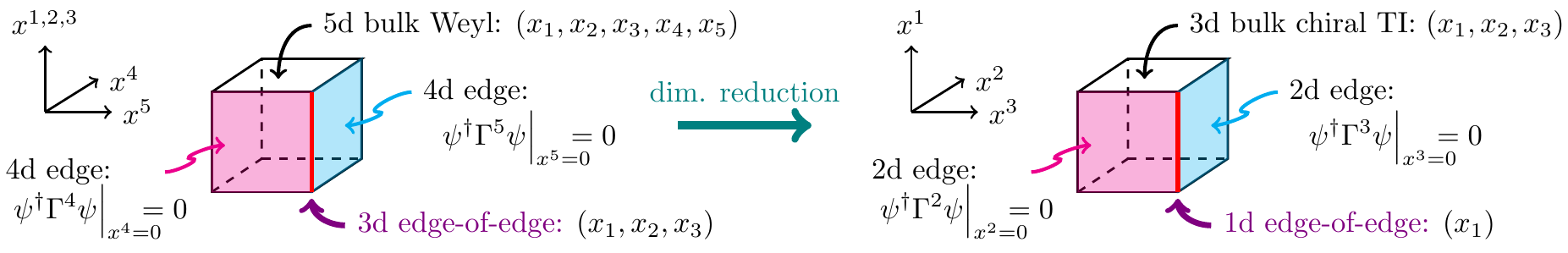}
	 \caption{The schematic picture of the corner state and the dimensional reduction from the 
	 5D Weyl semimetal to the 3D chiral topological insulator (class AIII), taken from FIG.~1. of \cite{hashimoto2017edge}.}
	 \label{two-b}
          \end{center}
\end{figure*}


The organization of this paper is as follows. 
In Sec.~\ref{revsec}, we briefly review the corner states in the continuum
5D Weyl semimetals, with the most generic boundary conditions,
obtained in \cite{hashimoto2017edge}. In Sec.~\ref{sec:rev}, we provide a relation between the popular HOTIs in the continuum
and the 5D Weyl semimetals. 
In Sec.~\ref{sec:hami}, the effective Hamiltonian of the edge state is shown to be identical to that of a 3D Weyl.
In Sec.~\ref{sec:top}, the topological charge carried by the most generic edge state of the 5D Weyl
is shown to be that of a Dirac monopole. 
Then in Sec.~\ref{sec:nest}, we explicitly calculate the nested Wilson loop of the 5D Weyl semimetal and show that
it is dictated by a monopole topological charge. The last section is for a summary and discussions. App.~\ref{sec:app}
is a proof of the most generic boundary condition of the 5D Weyl semimetal.
App.~\ref{sec:hermit} describes cases where corner states are absent in spite of the nontrivial topology.


\section{Review of corner states in 5D Weyl  semimetal}
\label{revsec}

In this section we give a review of generic corner (hinge) states  in 5D Weyl semimetals, 4-d class A
and 3D class AIII topological insulators,
obtained in \cite{hashimoto2017edge}.
The discovery of corner states \cite{hashimoto2017edge}
was independent of the discovery of multipole insulators \cite{benalcazar2017quantized}.
The corner states are called 
``edge-of-edge states" in \cite{hashimoto2017edge}, but in this paper we call them
corner states.\footnote{In all examples in this paper, corner states are localized at co-dimension two subspace,
so they are hinge states and the topology is second order. For simplicity, in this paper we call them corner states
and HOTIs.} See Fig.~\ref{two-b} (taken from \cite{hashimoto2017edge}) for the schematic view of the edge-of-edge states.

The Hamiltonian of the 5D Weyl semimetal \cite{lian2016five} in the continuum limit is given by
\begin{align}
H = \sum_{I=1}^5 p_I \Gamma_I
\label{5dH}
\end{align}
where $(p_1, \cdots, p_5)$ is the momentum in the 5D space, and $\Gamma_I$ is
the $4\times 4$ gamma matrices satisfying the Clifford algebra, $\{\Gamma_I, \Gamma_J\} = 2\delta_{IJ}$. 
It is the simplest to take the chiral representation of the gamma matrices,
\begin{align}
\Gamma_i \equiv \left(
\begin{array}{cc}
0 &-i\sigma_i \\
i \sigma_i & 0
\end{array}
\right) \quad (i=1,2,3) 
\nonumber 
\\
\Gamma_4 \equiv \left(
\begin{array}{cc}
0 &{\bf 1}_2 \\
{\bf 1}_2& 0
\end{array}
\right)\, ,
\quad  
\Gamma_5 \equiv \left(
\begin{array}{cc}
{\bf 1}_2&0 \\
0& -{\bf 1}_2
\end{array}
\right)\, . 
\end{align}
Here $\sigma_i (i=1,2,3)$ is the Pauli matrix.
The 4-d class A topological insulator is obtained by putting $p_5=m_5$ (constant) \cite{ryu2010topological}. 
\begin{align}
H_{\rm 4d \, A} = \sum_{i=1}^3p_i \Gamma_i + p_4 \Gamma_4 + m_5\Gamma_5  \, .
\label{H4A}
\end{align}
The 3D class AIII
topological insulator is a dimensional reduction $p_4=m$ and $p_5=0$,
\begin{align}
H_{\rm 3d \, AIII} = \sum_{i=1}^3p_i \Gamma_i + m \Gamma_4 \, .
\label{H3AIII}
\end{align}
The edge states of the 5D Weyl semimetals were studied in \cite{lian2016five}.

It was argued in \cite{hashimoto2017edge} that the most general boundary condition at $x_5=0$,
allowed by the Hermiticity of the system, is 
\begin{align}
\left({\bf 1}_2, -U_5^\dagger\right)\, \psi \biggm|_{x_5=0} = 0 \, .
\label{gU}
\end{align}
Here $U_5$ is a generic U(2) constant matrix. 
We will provide a rigorous proof of this in Appendix \ref{sec:app}. 
Even under a dimensional reduction to HOTI or other topological insulators, 
this generic boundary condition \eqref{gU} is not modified. Of course, the dimensional reduction 
should not be made along the direction perpendicular to the introduced surface (which is $x_5$ in
the example above).
Note that this boundary condition is the most general one which respects 
the Hermiticity of the system with the Hamiltonian \eqref{5dH} in the continuum limit. 
The genericity means that any microscopic boundary condition at $x_5=0$ which does not break the Hermiticity 
should reduce to a certain $U_5$ if the Hamiltonian in the continuum limit is given by \eqref{5dH}.  
Reverse engineering of finding a lattice boundary condition from a given $U_5$ is a nontrivial problem.

For surfaces not in $x_5=0$, one just makes a rotation in the 5D space together with the Gamma matrices.
For example, a surface at $x_4=0$ allows the following generic boundary condition \cite{hashimoto2017edge}, 
\begin{align}
\left( \frac12 (U_4^\dagger - U_4), {\bf 1}_2-\frac12 (U_4+U_4^\dagger)\right)\, \psi \biggm|_{x_4=0} = 0 \, .
\label{gU4}
\end{align}
Here, again, $U_4$ is a general constant U(2) matrix.

In \cite{hashimoto2017edge} it was shown that in a 5D Weyl semimetal \eqref{5dH} 
with two surfaces with most generic
boundary conditions \eqref{gU} and \eqref{gU4}, there exists generically a corner state
localized at $x_4=x_5=0$, with an energy $\epsilon$ given by a solution of the equation
\begin{align}
A\epsilon^2 -2 B \epsilon + C = 0
\end{align}
where
\begin{align}
&A \equiv 1-\cos^2\theta_4 \cos^2\theta_5 \, , 
\nonumber
\\
& B\equiv a_i p_i \cos\theta_5 \sin^2\theta_4 + b_i p_i \cos\theta_4\sin^2\theta_5 \, ,
\\
&
C\equiv
(a_i p_i)^2 \sin^2\theta_4 + (b_i p_i)^2 \sin^2\theta_5
-p_i^2 \sin^2\theta_5 \sin^2 \theta_4 \, .
\nonumber
\end{align}
In this expression we have parameterized the boundary condition parameters $U_5$ and $U_4$ as
\begin{align}
U_5 \equiv e^{i\theta_5}(a_0 {\bf 1}_2 + i a_i \sigma_i) \, , \quad
U_4 \equiv e^{i\theta_4}(b_0 {\bf 1}_2 + i b_i \sigma_i) \, , 
\end{align}
with $a_0^2+a_i^2 = b_0^2+b_i^2 = 1$.

The simplest example presented in \cite{hashimoto2017edge} was a corner state with a dispersion
$\epsilon = - p_1$, with the following boundary conditions at the intersecting surfaces,
\begin{align}
\left(\Gamma_4 - i \Gamma_3\right) \psi\biggm|_{x_4=0} = 0 \, ,
\quad
\left(\Gamma_5 - i \Gamma_2\right) \psi\biggm|_{x_5=0} = 0 \, .
\label{4352}
\end{align}
Another simple example in a class AIII topological insulator \eqref{H3AIII}
was given in \cite{hashimoto2017edge} as
a localized state at the corner $x_2=x_3=0$ with a dispersion $\epsilon=-p_1$ with the following boundary
conditions
\begin{align}
\left(\Gamma_2 - i \Gamma_5\right) \psi\biggm|_{x_2=0} = 0 \, ,
\quad
\left(\Gamma_3 - i \Gamma_4\right) \psi\biggm|_{x_3=0} = 0 \, .
\end{align}

The 5D Weyl semimetal \eqref{5dH} has a topological defect centered at $p_I=0$
from which a 1-form flux $*(F \wedge F)$ made of the second Chern class density emanates.
This can be easily confirmed by 
the dimensional reduction to 
the 4D topological insulator of class A \eqref{H4A}.
the bulk topology is the second Chern class of
the Berry connection \footnote{The slight difference from the renowned ADHM construction of 
instantons was studied in detail in \cite{hashimoto2016band}.}, 
\begin{align}
\frac{1}{16\pi^2}\int dp_1dp_2dp_3dp_4 \, {\rm tr}
\left[F_{\mu\nu} * F_{\mu\nu}\right] = -\frac{1}{2} {\rm sign}(m_5) \, .
\end{align}
The dimensional reduction corresponds to taking a slice of the originally 5D momentum space,
so the reduced system inherits partly the original topological structure.

Introducing a surface was interpreted as a T-duality in string theory \cite{hashimoto2016topological},
and because of that, the edge state on the surface of the 4D class A system
was shown to have an independent topological charge \cite{hashimoto2016topological}.
It leads to a prediction of having a corner (hinge) state once another surface is introduced to the system.
As a result, various corner (hinge) states were constructed in \cite{hashimoto2017edge}.


\section{5D Weyl and HOTI}
\label{sec:rev}

In this section we study the genericity of our 5D Weyl Hamiltonian, in particular in regards to HOTIs.
We specify the discrete symmetry which brings any four-band system to 5D Weyl semimetals.
If we regard some of the components of the 5-momentum $p_I$ as a constant or zero, then this system 
reduces to lower-dimensional semimetals/insulators. 
We describe dimensional reduction to two popular examples of higher-order topological insulators
provided in  \cite{benalcazar2017quantized,schindler2018higher}.

First, let us point out an important discrete symmetry which 
protects the Weyl Hamiltonian \eqref{5dH}.
Suppose we are interested in four-band material, then the Hamiltonian is a four-by-four Hermitian matrix.
Any such matrix $H$ can be expanded as
\begin{align}
H = p_0 {\bf 1}_4 + \sum_{I=1}^5 p_I \Gamma_I + i \sum_{I>J}p_{IJ}\Gamma_I\Gamma_J \, .
\label{Htry}
\end{align}
Here $p_0, p_I$ and $p_{IJ}$ are real parameters. 

In even dimensions $D=2n$, the $2^n$-dimensional 
representation of Clifford algebra is known to be unique, so the representation 
$\Gamma_I$ $(I=1,\cdots,2n)$
should be related to its transpose $(\Gamma_I)^T$ by a similarity transformation $C \Gamma_I C^{-1} = c (\Gamma_I)^T$,
with a constant $c=\pm 1$. 
Our 5 dimensional Gamma matrices are obtained by adding $\Gamma_5$, 
which is given by a product $\Gamma_1\Gamma_2\Gamma_3\Gamma_4$ up to a sign, 
to the $D=4$ case, 
so $\Gamma_5$ is subject to the same similarity transformation, and $c$ is known to be equal to 1.

Suppose we impose the invariance of the matrix $H$ \eqref{Htry} under the similarity transformation $C$,
\begin{align}
C H C^{-1} =  (H)^T \, .
\label{HT}
\end{align}
Then
we find that the last term $\Gamma_I \Gamma_J$ is not invariant under this transformation, thus the symmetric Hamiltonian
under $C$ transformation is
\begin{align}
H = p_0 {\bf 1}_4 + \sum_{I=1}^5 p_I \Gamma_I  \, .
\label{Htry2}
\end{align}
This is the Weyl Hamiltonian \eqref{5dH}, as the first term ${\bf 1}_4$ is a trivial addition to determine the zero of the 
whole energy spectrum. So we conclude that the similarity transformation $C$ brings the general four-band system to
the 5D Weyl semimetal. The resultant symmetric system has only five 
constant parameters, $p_I$ $(I=1,2,3,4,5)$.

The physical understanding of the similarity transformation 
\eqref{HT} is as follows. Our Gamma matrices are Hermitian, so we may replace \eqref{HT}
by
\begin{align}
C H C^{-1} = H^* \, .
\end{align}
This is known to be the PT symmetry (Parity and Time reversal symmetry) \footnote{We would like to thank R.~Okugawa for 
his suggestion.}
if we regard the coefficients $p_I$, $p_{IJ}$ do not change under the symmetry. 
In fact, if the coefficients are purely written by momenta, since the PT symmetry leaves momenta invariant, 
the coefficients are invariant. So in general the PT symmetry is our similarity transformation $C$.

The importance of the PT symmetry to restrict the structure of the four-band material is well-known \cite{yang2014classification}, 
in particular
for classifying Dirac semimetals. Dirac semimetals in the continuum limit is a particular dimensional reduction of
the 5D Weyl semimetal.
Very recently, it was reported \cite{wieder2020strong} that some Dirac semimetals host a higher-order topology,
which is along our claim that the 5D Weyl semimetal is 
a universal origin of a class of the higher-order topology.

As a consequence of the similarity $C$, the resultant spectra is doubly degenerate. This is a consequence of
the Kramers degeneracy in terms of the PT symmetry. As we will see in later sections, 
the degeneracy makes the Berry connection
to be non-Abelian, and the non-Abelian structure is necessary for the system to host higher-order topology \footnote{We would like to thank Y.~Hatsugai for clarifying this point.}.

Now, let us describe two popular examples of HOTI \cite{benalcazar2017quantized,schindler2018higher}
and show that these are dimensional reduction of 5D Weyl semimetals.

The first example is 
the famous 2D quadrupole insulator given
by Benalcazar {\it et al.} \cite{benalcazar2017quantized}. It has a lattice Hamiltonian
\begin{align}
H = & -\sin k_x \Gamma_3 + (-1+m_x + \cos k_x) \Gamma_4
\nonumber \\
& -\sin k_y \Gamma_1 - (-1+m_y+\cos k_y)\Gamma_2 \, ,
\end{align}
where $(k_x,k_y)$ is the 2D momentum and $m_x, m_y$ are nonzero constant. 
The expansion around $k_x, k_y, m_x, m_y \sim 0$ provides 
a continuum limit, 
\begin{align}
H_{\rm HOTI} = (-k_y)\Gamma_1 + (-m_y)\Gamma_2 + (-k_x)\Gamma_3 + m_x\Gamma_4 \, .
\label{Bel}
\end{align}
This system hosts a corner state \cite{benalcazar2017quantized}. 
The identification with the 5D Weyl semimetal is obvious,
\begin{align}
& p_1 = -k_y\, , \quad p_2 = -m_y\, , \quad
p_3 = -k_x\, , \quad p_4 = m_x\, , \nonumber\\
& p_5=0 \, .
\label{quad}
\end{align}
So, the $x$ direction in \cite{benalcazar2017quantized} is the $(-x_3)$ direction, and 
the $y$ direction  in \cite{benalcazar2017quantized} is the $(-x_1)$ direction.
This identification is made explicit to match the notation of \cite{benalcazar2017quantized}, but identification
with any $SO(5)$-rotated
frame leads to the same physical spectra.
In \cite{benalcazar2017quantized} the boundary conditions at the two orthogonally intersecting edges 
were 
derived (see the supplement Sec.~II of \cite{benalcazar2017quantized}),
and in our notation they are 
\begin{align}
\left({\bf 1}_4 - i \Gamma_3\Gamma_4\right)\psi\biggm|_{x_3=0} = 0 \, , \quad
\left({\bf 1}_4 + i \Gamma_1\Gamma_2\right)\psi\biggm|_{x_1=0} = 0 \, .
\label{3412}
\end{align}
These boundary conditions turn out to be identical to the ones \eqref{4352} given in \cite{hashimoto2017edge}. 
In fact, if we make the following renaming of coordinates,
\begin{align}
(x_1,x_2,x_3,x_4,x_5) \to
(x_5,x_2,x_4,x_3,-x_1),
\end{align}
this brings \eqref{4352} to \eqref{3412}.

Another popular example of the higher-order topological insulator was given by Schindler {\it et al.} in 
\cite{schindler2018higher}, a 3D insulator with $C^z_4 T$ symmetry,
\begin{align}
H = & \left(2+\sum_i \cos k_i\right) \Gamma_5
+ \sum_i \sin k_i \Gamma_i \nonumber \\
&+ \Delta_2 \left(\cos k_x - \cos k_y \right) \Gamma_4 \, .
\end{align}
Expanding this lattice Hamiltonian around the band inversion point $k_i = (\pi,\pi,\pi)_i + \tilde{k}_i$, a continuum Hamiltonian is
obtained as 
\begin{align}
H = \sum_i \tilde{k}_i \Gamma_i - \Gamma_5 \, .
\end{align}
This can be interpreted as the 3D class AIII Hamiltonian \eqref{H3AIII} under swapping $\Gamma_4$ and $\Gamma_5$, 
so it is 
given by a dimensional reduction from the 5D Weyl semimetal. The result of \cite{schindler2018higher} showing the hinge states
is consistent with \cite{hashimoto2017edge} claiming general hinge states for the class AIII topological insulators.
It should be noted here also that the HOTI \eqref{Bel} is a trivial dimensional reduction of 
a 3D class AIII topological insulator \eqref{H3AIII}.

Note that all known examples of HOTI use Hamiltonians with at least four bands \footnote{Chiral-symmetric HOTI \cite{okugawa2019second} is also of this type.}. 
It was shown in \cite{hashimoto2017edge} 
that 3D Weyl semimetals cannot host a corner state, meaning that two-band Hamiltonian cannot be a HOTI. 
From the generic behavior of the 5D Weyl semimetal hosting corner states, we may argue that among all possible four-band
Hamiltonians the structure of the gamma matrices is important to host the corner states. 
A physical reason for this is the edge topological charge \cite{hashimoto2016topological}, 
derived through an analogy to the Nahm construction
of monopoles and string T-duality.

\section{Edge Hamiltonian}
\label{sec:hami}

In this section we calculate the edge Hamiltonian for edge states in the 5D Weyl semimetal
with the generic boundary condition \eqref{gU}. We show explicitly that the edge Hamiltonian has
the structure of a 3D Weyl semimetal.

For the general boundary condition \eqref{gU}, 
the edge state wave function is expressed as \cite{hashimoto2017edge}: 
\begin{align}
\psi = 
\left(
\begin{array}{c}
{\bf 1}_2 \\
U
\end{array}
\right)
\xi \exp [i p_I x^I]\, . 
\label{est}
\end{align}
The edge state is given by \eqref{est}, and if we choose the basis 2-spinor as $\xi_1 = (1,0)^{\rm T}$
and $\xi_2 = (0,1)^{\rm T}$, then these $\xi_a$ ($a=1,2$) give the following basis for the 4-spinor,
\begin{align}
\psi_a \equiv 
\left(
\begin{array}{c}
{\bf 1}_2 \\
U
\end{array}
\right)
\xi_a \, .
\label{est2}
\end{align}
Using this, the effective Hamiltonian for the edge states,
\begin{align}
H^{\rm (eff)}_{ab} \equiv \int d^5 x\, \psi_a^\dagger H \psi_b \, , 
\end{align}
is calculated as 
\begin{align}
H^{\rm (eff)} = \cos\theta \, \tilde{p}_4{\bf 1}_2 + \sin\theta \, \tilde{p}_i\sigma_i \, , 
\label{edgeH}
\end{align}
where we decomposed the $U(2)$ matrix to a product of a $U(1)$ phase
$\theta$ and an $SU(2)$ matrix $U'$, 
\begin{align}
U = e^{i\theta} U'
\end{align}
and we defined the $SO(4)$ rotated momentum frame $(\tilde{p}_i, \tilde{p}_4)$ by
the following $SU(2)$ matrix relation
\begin{align}
(-i p_i\sigma_i + p_4) U' = -i \tilde{p}_i\sigma_i  + \tilde{p}_4 \, .
\end{align}
This edge Hamiltonian $H^{\rm (eff)}$ \eqref{edgeH} is of the form of a 3D Weyl semimetal, so the edge state is topological. 
Therefore, when another surface perpendicular to the original surface $x^5=0$ is introduced, 
a corner state is expected to exist at the intersection of the surfaces, 
due to the topological structure of the edge Hamiltonian.

Note that this argument is to present a general framework of the emergence of the corner states.
Whether the corner state actually exists or not in a particular given setup depends on more details of the system.
There are at least two reasons for this. 
First, if the effective Hamiltonian \eqref{edgeH} is that of a 3D Type II Weyl semimetal \cite{soluyanov2015type},
because the existence of an edge state for a given Type II Weyl semimetal depends on the 
surface direction \cite{hashimoto2019escape}, our 5D system may not give a corner state. 
Second, it is possible that the 3D Weyl Hamiltonian structure 
may be accidentally inactive for particular dimensional reductions to realistic dimensions.
For example, if the directions $\tilde{p}_i$ in \eqref{edgeH} are all dimensionally reduced, the Hamiltonian simply
reduces to a unit matrix which does not have any topological structure. 
Concerning the first reason, we describe a concrete example in App.~\ref{sec:hermit}.

We will calculate the topological charge in the next section. 
For the calculation we need the wave function which 
is defined by $\xi$. The effective Hamiltonian is diagonalized by a solution of
\begin{align}
\left(\pm\sqrt{\tilde{p}_i^2} {\bf 1}_2 - \tilde{p}_i \sigma_i \right) \xi = 0 \, .
\label{pmxi}
\end{align}
The energy $\epsilon$ is given by
\begin{align}
\epsilon = \tilde{p}_4 \cos\theta \pm \sqrt{\tilde{p}_i^2} \sin\theta \, ,
\end{align}
which is consistent with the result in \cite{hashimoto2017edge}. 

In obtaining the effective Hamiltonian, 
we didn't include the $x_5$ dependence in the wave function \eqref{est2}.
It is not a problem, because 
the relevant term disappears when evaluating the effective Hamiltonian, 
due to the Hermiticity condition \eqref{Herm}.
In order to extract the information about $x_5$ dependence from this effective Hamiltonian approach,
we instead consider the following equation
\begin{align}
H \psi_a = E_a \psi_a + \sum_{b=1,2} \beta_{ab} \tilde{\psi}_b
\label{HE}
\end{align}
where $\tilde{\psi}_b$ ($b=1,2$) is the basis 4-spinor spanning the space orthogonal 
to the edge state space spanned by $\psi_a$,
\begin{align}
\tilde\psi_a = 
\left(
\begin{array}{c}
{\bf 1}_2 \\
-U
\end{array}
\right) 
\xi_a \exp[ip_I x^I]
\, .
\end{align}
For the equation \eqref{HE} to be a Hamiltonian eigenequation for the edge state, we need to require
$\beta_{ab}=0$. Since $\beta$ is given by
\begin{align}
\beta_{ba} = \int d^5 x\, \tilde{\psi}_b^\dagger H \psi_a, 
\end{align}
we can calculate it explicitly as 
\begin{align}
\beta &
= 
\left(
\begin{array}{c}
{\bf 1}_2 \\
-U
\end{array}
\right)^{\!\!\dagger} 
H
\left(
\begin{array}{c}
{\bf 1}_2 \\
U
\end{array}
\right) 
\nonumber \\
&
= p_5 {\bf 1}_2 + i\left(
\sin\theta \, \tilde{p}_4  {\bf 1}_2
- \cos\theta \, \tilde{p}_i \sigma_i \right)\, .
\end{align}
Because the eigenvalue of this expression needs to vanish, we find that the momentum $p_5$
has to be pure imaginary and is given by
\begin{align}
p_5 = i \alpha \equiv -i \left(
\sin\theta \, \tilde{p}_4
\mp \cos\theta \, \sqrt{\tilde{p}_i^2}\right).
\end{align}
The signs are for the two edge states, respectively.
The edge states are localized at the surface $x_5=0$, 
thus the wave function should have a pure imaginary
momentum $p_5$. The normalizability of the wave function restricts the
possible momentum region of $(p_i, p_4)$ for the existence of the edge state.

\begin{figure*}[t]
	\begin{center}
	 \includegraphics[width=40em]{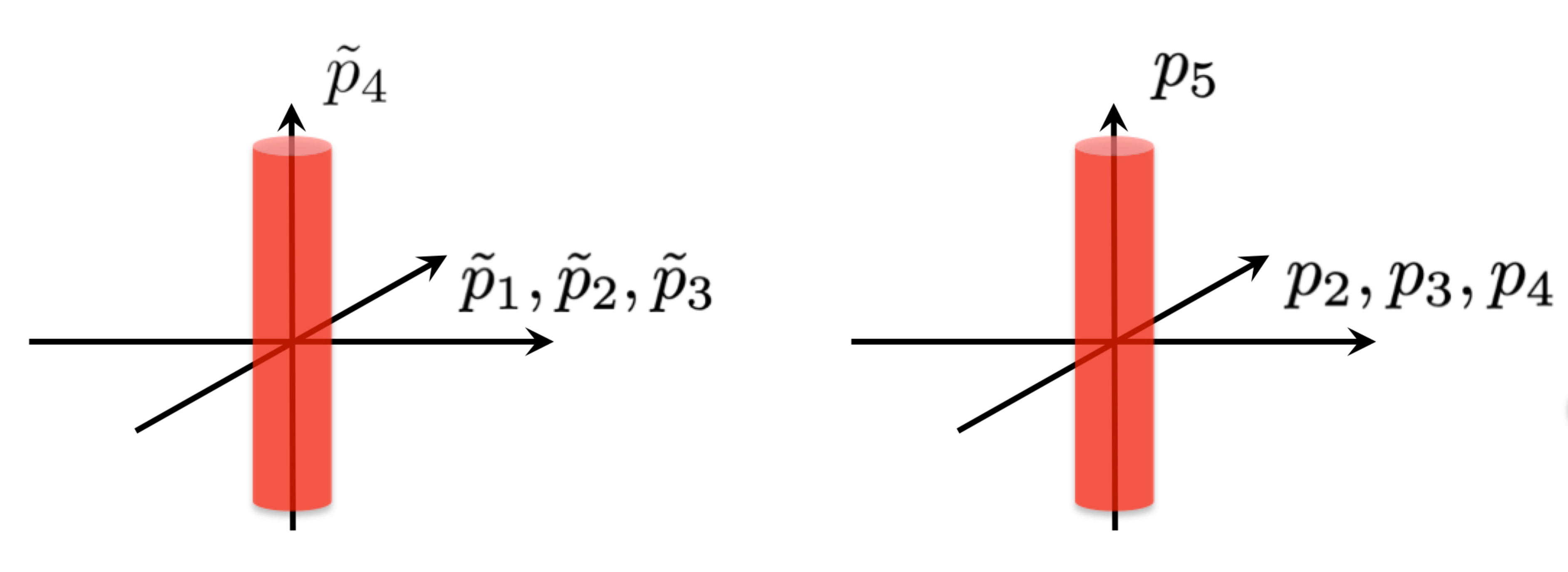}
	 \caption{
	 The structure of the topological charges.
	 Left: the topological charge distribution of the edge state. 
	 It is a Dirac monopole situated at the origin 
	 of $\tilde{p}_1, \tilde{p}_2, \tilde{p}_3$ space. They are aligned along the $\tilde{p}_4$ direction, 
	 so it is a monopole-string.
	 The 4-momentum $\tilde p_1,\cdots,\tilde p_4$ is rotated from the original 4-momentum, 
	 $p_1,\cdots,p_4$ for the edge at $x_5=0$, and 
	 $\tilde p_4$ direction in the original momentum space depends on the boundary condition. 
	 Right: the topological charge of the Hamiltonian $H_1$ of the nested Wilson loop along $p_1$, given in \eqref{H1result}. 
	 The Dirac monopole is located at the origin of 
	 $p_2,p_3,p_4$ space. It has no dependence in $p_5$, so the monopole string extends along the $p_5$ axis.
	 The direction of the monopole string depend on the gauge condition of the original 5D Berry connection, 
	 but independent of the gauge condition for the Berry connection for $H_1$. 
	 }
	 \label{fig:monopole}
          \end{center}
\end{figure*}

\section{Edge topological charge}
\label{sec:top}

In this section, we show that the topological charge of the generic edge state of the 5D Weyl semimetal
is a Dirac monopole. As we described in the introduction, the corner states in the 5D Weyl semimetals
are a consequence of the topological charge of the edge states. In \cite{hashimoto2016topological},
for a particular boundary condition of a 4D class A topological insulator, the edge state is shown to carry
a topological charge of a Dirac / nonAbelian monopole. It was argued in \cite{hashimoto2017edge}
that the edge topological charge can be generally the same for the 5D Weyl semimetals.
Here we explicitly confirm it, by providing the topological charge of the edge state of the 5D Weyl semimetal
\eqref{5dH}
with arbitrary boundary condition \eqref{gU}.

Let us calculate the topological charge of the edge state.
It was shown in \cite{hashimoto2017edge} that the edge state has a topological charge
which is the same type as that of the 3D Weyl semimetal. Here we demonstrate it explicitly.
For a particular choice of the boundary condition, the topological charge was calculated
in \cite{hashimoto2016topological}.

As we have seen in the previous section, the effective Hamiltonian for edge states 
is diagonalized by solutions of \eqref{pmxi}. 
We name the two solutions of \eqref{pmxi} as $\xi_+$ and $\xi_-$ depending on the $\pm$ sign
in \eqref{pmxi}. As they can be regarded as the energy eigenstates of 
a 3D Weyl semimetal Hamiltonian \eqref{edgeH}, 
it naturally leads us to suggest that 
the edge state has a topological charge. In fact, we can calculate the Berry connection
associated with $\xi_+$ and show that it is the connection for a Dirac monopole, 
\begin{align}
\tilde{A}_1 + i \tilde{A}_2 = \frac{ i(\tilde{p}_1 + i \tilde{p}_2)}{2 \sqrt{\tilde{p}_i^2} (\sqrt{\tilde{p}_i^2}-\tilde{p}_3
)} \, , \quad \tilde{A}_3=\tilde{A}_4=0 \, .
\label{mono}
\end{align}
Here $\tilde{A}$ is the connection in the rotated frame spanned by $(\tilde{p}_i, \tilde{p}_4)$.
It is defined by\footnote{
The definition needs the integration over $dx_5$ (along the direction perpendicular to the surface),
which was suggested by the Nahm construction of monopoles \cite{hashimoto2016topological}.
}
\begin{align}
\tilde{A}_{i} =  
\int d^5 x\, i\psi^\dagger \frac{d}{d\tilde{p}_{i}}\psi \, , \quad
\tilde{A}_{4} = 
\int d^5 x\, i\psi^\dagger \frac{d}{d\tilde{p}_{4}}\psi \, .
\end{align}

The Dirac monopole \eqref{mono} sits at the origin of $(\tilde{p}_i)$ ($i=1,2,3$) space, and 
it extends to the $\tilde{p}_4$ direction. So this is a monopole-string in the 4D momentum space.
The location of the monopole-string is at $\tilde{p}_i=0$. (Note that the ``string" of the ``monopole-string"
is different from the Dirac string associated with a 3-dimensional monopole; the former is physically extending
along the $\tilde{p}_4$ direction while the latter is a gauge artifact.) 
See Fig.~\ref{fig:monopole} Left for 
the configuration of the monopole-string.

If we go back to the original momentum space spanned by $(p_i,p_4)$, this monopole-string is also 
rotated back. The direction of the extension of the monopole-string is specified by a straight line
\begin{align}
{\rm tr}
\left[ \sigma_j(-i p_i\sigma_i + p_4) U'\right] = 0 \,  \quad (j=1,2,3)
\end{align}

Let us study how the topological charge looks like after the dimensional reduction to the HOTI \eqref{Bel}.
In the notation above, the surface is introduced at $x_5=0$, so the reduction to \eqref{Bel} means 
placing three of the other momenta by some constant, for example
\begin{align}
p_2=m_x\, , \quad p_3 = m_y\, , \quad  p_4=0 \, .
\label{slicehoti}
\end{align}
Upon this reduction we have a 2D system spanned by $p_1$ and $p_5$.
The edge topological charge is seen only as a slice \eqref{slicehoti} of the monopole-string.
Only possible index associated with the edge state of the dimensionally
reduced system is the Wilson loop, 
\begin{align}
W \equiv \int_{-\infty}^{\infty} dp_1 \, A_1 \, ,
\end{align}
since the valid quantum number of the edge state is only $p_1$.
This $W$ is defined as an integral over the slice.
It is easy to see that this quantity is generically nonzero, due to the monopole-string in
the higher-dimensional momentum space. However, the value of $W$ is not quantized, and 
it depends on the parameters $m_x$, $m_y$ and $U$. In particular, it depends on
the angle between the edge momentum direction (the $p_1$ axis) and the plane
spanned by the $\tilde{p}_4$ axis and the $\tilde{p}_3$ axis. When they are parallel to each other,
$W$ vanishes, due to the vanishing components of \eqref{mono}.
This complication always emerge after dimensional reductions, because the topological charge
defined in the higher dimensional momentum space can never be captured naturally just by looking at the connections
on the slices.


\section{Nested Wilson loop}
\label{sec:nest}

So far, we have presented the topological structure of the most generic edge states of the 
continuum 5D Weyl semimetal. On the other hand, 
it was shown that the nested Wilson loop \cite{benalcazar2017quantized}
captures the topological property of HOTIs\footnote{Entanglement polarization
\cite{fukui2018entanglement} is one of the other quantities to characterize the higher-order topology from
the bulk wave function. In \cite{fukui2018entanglement} an explicit entanglement polarization formula
which is applicable also to the 5D Weyl semimetal was provided, and it is of the form of a Dirac monopole.
So it shares the same topological property.}, so here we explicitly 
calculate the nested Wilson loop for the continuum 5D Weyl semimetal.
We show that the nested Wilson loop is identical to a Wilson loop in a 4D momentum space with 
a Dirac monopole-string.

First we obtain the normalized energy eigenstates of the bulk Hamiltonian \eqref{5dH}.
The energy eigenvalues are $\pm p$ with $p\equiv\sqrt{p_I^2}$, and we take the positive energy eigenstates.
The degenerate two eigenstates are aligned together to form a $4\times 2$ matrix $\Psi$
satisfying $H \Psi = p \Psi$,  
\begin{align}
\Psi =  \sqrt{\frac{p-p_5}{2p}}
\left(
\begin{array}{c}
{\bf 1}_2 \\
F
\end{array}
\right) 
\label{PsiF}
\end{align}
where
\begin{align}
F \equiv \frac{1}{p_5-p}(p_4 {\bf 1}_2 + i p_i \sigma_i) \, .
\end{align}
With this energy eigenstate, the Berry connection 
\begin{align}
A_I \equiv i \Psi^\dagger \frac{\partial}{\partial p_I} \Psi
\end{align}
can be calculated explicitly, 
\begin{align}
& A_1 = C_A \left(
p_4 \sigma_1 + p_3 \sigma_2-p_2 \sigma_3
\right)
 \, , 
 \label{A1B}\\
& A_2 = C_A \left(
p_3 \sigma_1 - p_4 \sigma_2+p_1 \sigma_3
\right)
 \, , \\
& A_3 = C_A \left(
-p_2 \sigma_1 - p_1 \sigma_2-p_4 \sigma_3
\right)
 \, , \\
& A_4 = C_A \left(
-p_1 \sigma_1 + p_2 \sigma_2+p_3 \sigma_3
\right)
 \, , 
 \label{A4B}\\
& A_5=0 \, ,
\label{A5}
\end{align}
with an overall common coefficient $C_A \equiv 1/(2p(p+p_5))$.

Let us calculate the Wilson loop (line) of the Berry connection along the $p_1$ direction,
\begin{align}
W_1 \equiv {\rm P} \exp \left[ i \int_{-\infty}^\infty A_1 dp_1\right] \, 
\end{align}
where ``P" means a path-ordering which makes $W$ gauge-covariant.
We diagonalize the Berry connection $A_1$ using a unitary matrix $V_1$,
\begin{align}
A_1 = \lambda V_1^\dagger \sigma_3 V_1. 
\end{align}
Here $\pm \lambda$ is the eigenvalue of $A_1$, 
\begin{align}
\lambda \equiv \frac{\sqrt{p_2^2+p_3^2+p_4^2}}{2p(p+p_5)} \, .
\end{align}
Since the unitary matrix $V_1$ does not depend on $p_1$, the path-ordering becomes
trivial and we have
\begin{align}
W_1 = V_1^\dagger
\exp \left[ i \sigma_3 \int_{-\infty}^\infty \lambda dp_1\right] V_1\, . 
\end{align}
The integral is evaluated explicitly and we obtain
\begin{align}
W_1 = V_1^\dagger \left(
\begin{array}{cc}
e^{is} & 0 \\
0 & e^{-is}
\end{array}
\right)
V_1
\end{align}
with\footnote{
It is amusing to point out that this expression of $s$ is quite similar to the Chern-Simons 3-form
of the 3D class AIII system, see eq.~(94) of \cite{ryu2010topological}.
}
\begin{align}
s \equiv 2 \arctan \sqrt{\frac{\sqrt{p^2-p_1^2}-p_5}{\sqrt{p^2-p_1^2}+p_5}}\, .
\end{align}
To evaluate the nested Wilson loop, we consider a Wilson loop Hamiltonian
\begin{align}
H_1 \equiv -i \log W_1
\end{align}
which is calculated as
\begin{align}
H_1 = V_1^\dagger \sigma_3 V_1 s \, .
\label{H1result}
\end{align}
This $H_1$ is interpreted as some sort of the edge Hamiltonian for a 4D system 
on the edge at $x_1 = \text{const.}$ surface as 
it is obtained from the Wilson loop along the $p_1$ direction. 
This $H_1$ depends on the gauge condition, and in the present gauge, it
is independent of $p_5$ apart from the overall factor $s$. 

Now, we consider the eigenstates of $H_1$. 
Since the matrix part of this $H_1$ is proportional to $A_1$ itself,
and it is of the form of a 3D Weyl semimetal, 
the Berry connection of the eigenstate of this $H_1$ is that of a Dirac monopole
located at $p_2=p_3=p_4=0$. The monopole extends along $p_5$, so, precisely speaking,
the topological defect is a monopole-string. 
The direction depends on the gauge condition of the 5D Berry connection \eqref{A1B}--\eqref{A5}, 
but independent of the gauge condition of the Berry connection of the eigenstate of $H_1$. 
See Fig.~\ref{fig:monopole} Right for 
the configuration of the monopole-string.

Therefore, the nested Wilson loop is identical to a Wilson loop of a Dirac monopole-string.
For example, a nested Wilson loop along the $p_4$ direction at $p_2=p_3=-\infty$ is calculated as
\begin{align}
W^{(1)}_4 = \frac{\pi}{2} \, .
\end{align}
Here ``$W^{(1)}_4$" means the nested Wilson loop along $p_4$ obtained through the Wilson loop 
along $p_1$.

It is easy to evaluate other nested Wilson loops, $W^{(I)}_J$.
All nested Wilson loops are nontrivial due to the topological structure of the monopole-string,
except for the ones involving the $p_5$ direction $I=5$ (or $J=5$). 
Since in our gauge condition the $p_5$ direction is special
as the Berry connection vanishes there (see \eqref{A5}), the nontrivial structure is hidden
for Wilson loops associated with $p_5$. 

Note that the trivialization of the path-ordering is not accidental; it is due to a careful choice
of the gauge condition of the Berry connection, which originates in the choice of the basis of 
the eigen states \eqref{PsiF}. In fact, we could have chosen some other set of basis of the degenerate 
eigenstates, for example \eqref{PsiF}
multiplied by some arbitrary $U(2)$ unitary matrix from the right, without losing the orthonormality. 
This matrix can depend on $p_I$, and this results in a gauge transformation
of the Berry connection. So the choice of the gauge condition given by \eqref{PsiF} makes the calculation of
the Wilson loop possible.\footnote{The direction of the monopole-string for the nested Wilson loop
depends on the gauge condition of the
5D Berry connection. On the other hand, in Section \ref{sec:top} the direction of the monopole-string of the 
edge topological charge depends on the boundary condition. To clarify any possible relation among these two
would be an interesting future problem.}

It is amusing to point out that the calculation of the nested Wilson loop Hamiltonian
$H_1$ is almost identical to what Atiyah and Manton did in 1989 \cite{Atiyah:1989dq}, so-called Atiyah-Manton approach for
Skyrmions.
The Hamiltonian $H_1$ corresponds to the pion field of the Skyrme model 
\cite{Skyrme:1961vq,Skyrme:1961vr,Skyrme:1962vh}.
The wrapping number, which is the baryon number in the Skyrme model, is nontrivial 
when the pion field has the hedge-hog ansatz. This hedge-hog ansatz is nothing but the
3d Weyl semimetal Hamiltonian in condensed matter physics. 
In this analogy to make sense, we need to exchange $x$ in the Skyrme model 
and $p$ in condensed matter physics, with a careful check of
the Berry connection \eqref{A1B} -- \eqref{A4B} to completely coincide with a renowned BPST instanton \cite{Belavin:1975fg},
as demonstrated in \cite{hashimoto2016band}.

We conclude that the nested Wilson loops of the 5D Weyl semimetal at the continuum
are topologically nontrivial and the topological structure is that of a Wilson loop in 
a monopole-string. $W^{(I)}_J$ 
$(I\neq 5)$ is identical to a Wilson loop along $p_J$ of a monopole-string where the Dirac monopole 
center is located at $p_K=0$ $(K\neq I, K \neq 5)$.  

This picture survives any dimensional reduction along the directions $\neq p_I$ and $\neq p_J$. 
In particular, the nested Wilson loop $W^{(1)}_2$ is of course identical with that evaluated
in the quadrupole insulator \cite{benalcazar2017quantized} upon the dimensional reduction \eqref{quad}.


\section{Summary and discussion}

In summary, we have bridged the higher-order topology of the 5D Weyl semimetals \cite{hashimoto2017edge}
and that of HOTIs, by showing that their Hamiltonians are related each other in the continuum. We have found that 
the 5D Weyl semimetal has the topological structure of the Dirac monopole from every aspect: the edge Hamiltonian
(Sec.~\ref{sec:hami}), the edge topological charge (Sec.~\ref{sec:top}), 
the nested Wilson loop and the entanglement polarization (Sec.~\ref{sec:nest}).
Therefore, the 5D Weyl semimetal can be regarded as 
a universal origin of a class of HOTIs.
The 5D Weyl semimetal is obtained in the continuum limit of 
four-bands with two pairs of double degeneracy which is protected by the PT symmetry. 
Note that our argument is effective at the continuum limit where various crystalline symmetries disappear,
so our universality refers to 
that of a universality class in the continuum limit or in the low-energy limit 
which is not protected by crystalline symmetries. 
As a scope of this work, this offers a novel criterion for how one can obtain HOTIs, due to its continuum structure.

We have described relations between the continuum 5D Weyl semimetals and the popular HOTI models.
The corner states of the 5D Weyl semimetal, obtained in \cite{hashimoto2017edge}, are shown in various manner
to lead to genetic HOTIs, as originally anticipated in \cite{hashimoto2017edge}. We have explicitly calculated the
topological charge of the edge state of the continuum 
5D Weyl semimetal, with the most generic boundary condition on the surface,
and have shown that it is a Dirac monopole (Precisely, it is a 
monopole-string in a 4D momentum space). The effective Hamiltonian of the 
edge state is shown to possess the structure of a 3D Weyl semimetal, which is consistent with the topological charge.
These calculations generalize \cite{hashimoto2016topological} and 
confirm the edge topological structure of \cite{hashimoto2017edge}, leading to the generic
existence of the corner state. Furthermore, we have given 
the explicit calculation of the nested Wilson loop of the continuum 5D Weyl semimetal, and 
found that the topological 
structure is identical to that of a Wilson loop in the 4D momentum space with a Dirac
monopole-string. The 5D Weyl semimetals can be dimensionally reduced to the realistic
HOTI models, while keeping the topological structure originated in the five dimensions.
It is worth being emphasized that the three methods, the edge Hamiltonian, the edge topological charge,
and the nested Wilson loop, do not require how the two surfaces intersect. Hence the universality
is applied for intersection of surfaces at a generic angle.

Our study and \cite{hashimoto2017edge} suggests that one of the bulk origins 
of generic corner states is the second Chern class. 
In fact, the edge topology which is the Dirac monopole is a consequence of the T-duality of the bulk Yang-Mills instanton
hosting the second Chern class
\cite{hashimoto2016topological}, 
and the bulk 5D Weyl semimetal has the second Chern class on the four-dimensional slice \cite{hashimoto2016band}.
This structure may be hidden once a dimensional reduction to lower dimensions is made. However,
since there exists a generic mechanism of the existence of the corner states due to the 5D Weyl semimetals,
generically after the dimensional reduction the possibility of having the corner states remains.

To look back, 
the study \cite{hashimoto2017edge} was motivated by elementary particle theories which are often in the continuum limit.
Although detailed study in condensed matter physics requires lattice Hamiltonians, 
which are lacking in \cite{hashimoto2017edge}, 
continuum limit generically provides a universal viewpoint which is irrelevant to detailed microscopic theories.
Bridging particle physics and condensed matter physics has played quite important role so far.
In particle physics, it has been known for many years that the anomaly inflow argument \cite{callan1985anomalies} provides 
fermion modes localized at co-dimension two hypersurface (that is, a string in three spatial dimensions), and this is 
one historical origin of hinge states. Based on this,
fermion localization at the intersection of surfaces in six dimensions was studied \cite{Neuberger:2003yg,Fukaya:2016ofi} 
for an application to 4D chiral gauge theories. Further symmetry/anomaly arguments may lead to 
various interesting researches bridging the particle and condensed matter physics.
In particular, particle physics is formulated to treat intrinsically many-body systems, and a quantum-field-theoretic
definition of multipoles \cite{kang2019many} may be of importance.

Some comments on possible future directions are in order.
In this paper we concentrated on the second-order topological insulators and hinge states, and a
generalization to third-order (and higher) is important. To host the third order, one needs 8 bands,
so the instanton may be an octonionic instanton \cite{Fubini:1985jm}. Some usage of higher-dimensional
Clifford algebra \cite{Nakamula:2016srw,Nakamula:2016wwv} may help building the bridge.
It is expected that on the surface of the eight-band system the BPST instanton emerges as the
surface topological charge.

It should be noted that Weyl semimetals admit various deformations: the tilt of the Weyl cones, 
producing Type II Weyl semimetals \cite{soluyanov2015type},
as well as Type III and Type IV Weyl semimetals
\cite{nissinen2017type}. In particular, similarities to black hole spacetime was suggested 
\cite{volovik2017lifshitz}, and its relevance to the surface state was studied
\cite{hashimoto2019escape}. 
The tilt of the edge dispersion can be interpreted \cite{hashimoto2016band} as a monopole
in non-commutative space \cite{hashimoto1999monopoles,gross2001solitons}, and these ideas may be
unified in higher dimensions of momenta.
The localization of the fermions at corners may be affected by the deformation, 
as was reported for example in non-Hermitian deformations
\cite{liu2019second}.
A deformation which breaks the PT symmetry was analyzed in \cite{lian2016five}, and it would be interesting 
whether the corner states may survive or not.
Furthermore, the ``deformation'' going back to the microscopic Hamiltonian 
will necessarily introduce a Brillouin zone which suffers from the doubling of the Weyl points \cite{lian2017weyl,chen2019doubling}. 
The continuum limit corresponds to a zoom-in on a small region around a Weyl point. 
As long as the Weyl points of the opposite signature are separated,
our monopole-string structure will persist, and the corner state will survive.
As well as these possible deformations, 
one should be careful in treating the dimensional reduction, as various dimensional reduction provides deformations
which could change the topological properties. 
For example, a possible relation between the interpretation of HOTI as a piled-up
Chern insulators \cite{matsugatani2018connecting} and our dimensional reduction needs to be clarified.

\acknowledgments

K.H.~would like to thank T.~Neupert, Y.~Hatsugai, B.~Kang, G.~Y.~Cho, R.~Okugawa 
and H.~Watanabe for valuable discussions. 
He also likes to thank H.~Fukaya, M.~Koshino and M.~Oshikawa for valuable comments.
This work is supported in part by JSPS KAKENHI Grant No.~JP17H06462.

\appendix

\section{Generic boundary condition}
\label{sec:app}

It was studied in \cite{hashimoto2017edge} that all possible boundary conditions of the continuum 
5D Weyl semimetal
\eqref{5dH}
are parameterized by a $U(2)$ matrix (see \cite{hashimoto2017boundary} for the case of the 3D Weyl semimetal).
Here we derive the same result in a more rigorous manner.
We use this result in Sec.~\ref{sec:top} and Sec.~\ref{sec:hami} 
to evaluate the topological charge of the generic edge state and 
the effective edge Hamiltonian.

We put the boundary surface at $x^5=0$ without losing its generality. The boundary condition is of the form
\begin{align}
M \psi\biggm|_{x_5=0} = 0
\label{Mpsi}
\end{align}
where $M$ is some constant matrix, and $\psi$ is the wave function. Since this $M$ has to have four
linearly independent eigen vectors to constrain the whole wave function space, $M$ is diagonalizable.
And $M$ has to project out half of the degrees of freedom of $\psi$ at the boundary, 
so two of the four eigenvalues should be zero.
Therefore, there exists a regular matrix $B$ with which
\begin{align}
M = B^{-1} D B, \quad
D = {\rm diag} \, \left( 0,0,a,b\right)\, ,
\end{align}
with $a,b\neq 0$. Decomposing $\psi$ and $B$ as their 2-spinor subspaces,
\begin{align}
\psi \biggm|_{x_5=0} \equiv
\left(
\begin{array}{c}
\xi
\\
\eta
\end{array}
\right)\, ,
\quad
B =
\left(
\begin{array}{cc}
B_1&B_2 \\
B_3& B_4
\end{array}
\right)\, ,
\end{align}
we can rephrase \eqref{Mpsi} as
\begin{align}
B_3 \xi + B_4 \eta = 0 \, .
\label{b3b4}
\end{align}
When $\det B_4 \neq 0$, we solve this as 
\begin{align}
\eta = - B_4^{-1}B_3 \xi \, .
\label{B4B3}
\end{align}

Next, we impose the boundary Hermiticity condition: for any $\psi_i$ satisfying $M \psi_i=0$,
we need  
\begin{align}
\psi^\dagger_i \Gamma_5 \psi_j = 0 \, .
\label{Herm}
\end{align}
The latter equation is equivalent to
\begin{align}
\xi^\dagger_1 \xi_2 = \eta^\dagger_1 \eta_2 \, .
\label{xixietaeta}
\end{align}
substituting the solution \eqref{B4B3} of the boundary condition \eqref{Mpsi} to this equation, we find
\begin{align}
(B_4^{-1} B_3)^\dagger B_4^{-1}B_3 = {\bf 1}_2 \, ,
\end{align}
as \eqref{xixietaeta} needs to be satisfied for arbitrary $\psi$ satisfying the boundary condition \eqref{Mpsi}.
This shows that $-B_4^{-1} B_3$ is a $U(2)$ unitary matrix $U$, so we derive the generic boundary condition
\begin{align}
\eta = U \xi \, 
\label{gU2}
\end{align}
for a $U(2)$ matrix $U$.

There remains the case of $\det B_4 = 0$. In fact, it is possible to show that 
with $\det B_4 = 0$ the condition \eqref{xixietaeta} cannot be met except for a trivial solution $\eta=0$,
as in the following.
So $\det B_4\neq 0$ is necessary for \eqref{Mpsi} to be a consistent boundary condition. 

When $\det B_4=0$ and $B_4\neq 0$, we name the zero eigenvector as $\eta^{(0)}$. Choose a vector $\eta^{(1)}$ which is
perpendicular to $\eta^{(0)}$, as $(\eta^{(0)})^\dagger \eta^{(1)}=0$. 
\begin{itemize}
\item If $\det B_3 \neq 0$, the vectors which satisfy the boundary condition \eqref{b3b4} is 
\begin{align}
\psi = \left(\begin{array}{c} 0 \\ \eta^{(0)} \end{array}\right) \, , \quad 
\tilde{\psi} = \left(\begin{array}{c} B_3^{-1}B_4 \eta^{(1)} \\ \eta^{(1)} \end{array}\right) \, .
\end{align}
\item If $\det B_3=0$ and $B_3\neq 0$, 
choose the zero mode of $B_3$ as $\xi^{(0)}$, then we obtain the vectors satisfying \eqref{b3b4}
as
\begin{align}
\psi = \left(\begin{array}{c} 0 \\ \eta^{(0)} \end{array}\right) \, , \quad 
\tilde{\psi} = \left(\begin{array}{c} \xi^{(0)} \\ 0 \end{array}\right) \, .
\end{align}
\end{itemize}
If $B_4=0$, then use two zero-modes to form a solution to \eqref{b3b4} as
\begin{align}
\psi = \left(\begin{array}{c} 0 \\ \eta^{(0)} \end{array}\right) \, , \quad 
\tilde{\psi} = \left(\begin{array}{c} 0 \\ {\eta'}^{(0)} \end{array}\right) \, .
\end{align}
In all of these three cases, we construct generic vectors satisfying \eqref{b3b4} as
\begin{align}
\psi_i \equiv a_i \psi + b_i \tilde{\psi} \, ,
\end{align}
with arbitrary complex constants $a_i, b_i$. Then we substitute them to the Hermiticity condition \eqref{Herm}.
It turns out that for all of these three cases, \eqref{Herm} for arbitrary $a_i$ and $b_i$ leads to $\eta^{(0)}=0$.
Therefore there is no solution to the boundary condition when $\det B_4=0$.

\section{Hermiticity condition for edge Hamiltonian}
\label{sec:hermit}

In order to have a corner state, 
the edge effective Hamiltonian \eqref{edgeH} must obey the hermiticity condition 
at the corner of the edge. 
However, 
for a certain boundary condition of the edge,
the effective Hamiltonian \eqref{edgeH}
takes the form of that of the Type II Weyl semimetals, and 
depending on the direction of the other edge it may 
not host any corner state. 
Here, we will show the condition for which 
the 5D Weyl semimetal does not have the corner state. 

We consider the corner at $x^4 = 0$ on the edge at $x^5 = 0$. 
The rotated momentum frame is defined by using an $SU(2)$ matrix $U'$, 
which is expressed generically as 
\begin{equation}
 U' = \cos\phi \, \mathbf{1}_2 + i \sin\phi \, n_i \sigma_i \ , 
\end{equation}
where $n_i$ is a unit vector, which can be taken as 
$n_i = (0,0,1)_i$ without loss of generality. 
Then, the edge Hamiltonian is expressed as 
\begin{align}
 H^{\text{(eff)}} 
 &= 
 \cos \theta \left(\cos\phi \, p_4 + \sin\phi \, p_3 \right) 
\notag\\&\quad
 + \sin\theta \left[\tilde p_1 \sigma_1 + \tilde p_2 \sigma_2 
 +\left(\cos\phi \, p_3 - \sin\phi \, p_4\right) \sigma_3 \right] \ , 
\end{align}
where $\tilde p_1$ and $\tilde p_2$ are linear combinations of $p_1$ and $p_2$. 
As the momentum $p_4$ acts on the wave function $\psi$ as the derivative $- i \partial_4$, 
the hermiticity condition requires the surface term at the boundary (the corner at $x^4=0$) vanish, 
\begin{equation}
 \psi_1^\dag \left(\cos \theta \cos \phi \, \mathbf{1}_2 
 - \sin \theta \sin \phi \, \sigma_3\right) \psi_2 
 = 0 \, .
\end{equation}
This condition must be satisfied for arbitrary wave functions 
$\psi_1$ and $\psi_2$ which satisfy the boundary condition at the corner. 
For $\psi_1 = \psi_2 = (v_1,v_2)^T$, the condition gives 
\begin{equation}
 \cos \theta \cos \phi \left(|v_1|^2 + |v_2|^2\right) 
 - \sin \theta \sin \phi \left(|v_1|^2 - |v_2|^2\right) 
 = 0 \, , 
\end{equation}
which has no solution for 
\begin{equation}
 |\tan \theta \tan \phi | \, < 1 \ . 
 \label{NoCorner}
\end{equation}
Therefore, 
for the boundary condition with \eqref{NoCorner},
the 5D Weyl semimetal has no corner state at $x^4 = x^5 = 0$. 

The parameters $\theta$ and $\phi$ are the ones for the boundary condition at $x^5=0$.
This means that one can even control the existence of the corner state by just
modifying a boundary condition on a single surface. 
The corner state may become unstable under the adiabatic change of the boundary condition
parameters of the surface. Here, a peculiar interplay between the higher order topology and
the Type II Weyl semimetal is found.

\bibliography{edge2}

\end{document}